# NMR of $^{31}$P Nuclear Spin Singlet States in Organic Diphosphates

Stephen J. DeVience[a,1,*], Ronald L. Walsworth[b,c,2], Matthew S. Rosen[c,d]

a. Department of Chemistry and Chemical Biology, Harvard University, 12 Oxford St., Cambridge, MA 02138, USA.

b. Harvard-Smithsonian Center for Astrophysics, 60 Garden St., Cambridge, MA 02138, USA.

c. Department of Physics, Harvard University, 17 Oxford St., Cambridge, MA 02138, USA.

d. Athinoula A. Martinos Center for Biomedical Engineering, Massachusetts General Hospital, 149$^{th}$ Thirteenth St., Charlestown, MA 02129, USA

1. Present address: Scalar Magnetics, LLC, 3 Harolwood Ct., Apt C, Windsor Mill, MD 21244, USA.

2. Present address: Quantum Technology Center, University of Maryland, 8228 Paint Branch Dr., College Park, MD 20742, USA.

*Corresponding author. Address: Scalar Magnetics, LLC, 3 Harolwood Ct., Apt C, Windsor Mill, MD 21244, USA. stephen@scalarmag.com

Email addresses: stephen@scalarmag.com (S. J. DeVience), walsworth@umd.edu (R. L. Walsworth), mrosen@mgh.harvard.edu (M. S. Rosen)

**Abstract**

$^{31}$P NMR and MRI are commonly used to study organophosphates that are central to cellular energy metabolism. In some molecules of interest, such as adenosine diphosphate (ADP) and nicotinamide adenine dinucleotide (NAD), pairs of coupled $^{31}$P nuclei in the diphosphate moiety should enable the creation of nuclear spin singlet states, which may be long-lived and can be selectively detected via quantum filters. Here, we show that $^{31}$P singlet states can be created on ADP and NAD, but their lifetimes are shorter than $T_1$ and are strongly sensitive to pH. Nevertheless, the singlet states were used with a quantum filter to successfully isolate the $^{31}$P NMR spectra of those molecules from the adenosine triphosphate (ATP) background signal.

**Introduction**

Organophosphates play a critical role in biology as energy carriers. Adenosine triphosphate (ATP) is the main currency of energy for the cell and is central to cellular metabolism [1,2]. Energy is released and used to drive metabolic processes by breaking the phosphate-phosphate bond, producing adenosine diphosphate (ADP) and inorganic phosphate. At the same time, new ATP is created from ADP via glycolysis, the citric acid cycle, and the electron transport chain, all driven by the breakdown of sugars, fatty acids, and proteins. ATP concentration *in vivo* is typically several times that of ADP, and their ratio provides information about cellular energy status and mitochondrial function [3,4]. Another organophosphate, nicotinamide adenine dinucleotide (NAD), is a cofactor for many metabolic pathways in which it is converted between its oxidized form NAD$^+$ and its reduced form NADH. The ratio between NAD$^+$ and NADH reflects the oxidoreductive state of the cell [5], and the total NAD concentration can change as a result of aging and neurodegenerative disease [6].

*In vivo* $^{31}$P NMR spectroscopy and imaging are commonly used to quantify the relative concentrations of ATP, ADP, and NAD non-invasively. Applications include the assessment of traumatic brain injury, investigation of aging, and diagnosis of musculoskeletal diseases [7-11]. However, detection and quantitation of these molecules can be difficult due to spectral overlap, particularly in the -11 to -12 ppm spectral region where lines from ATP, ADP, and NAD all occur. While some lines can easily be resolved with high-field, high-resolution NMR spectroscopy [10,12], lower field strengths used for human MRI make this challenging; thus

ADP concentration is often calculated indirectly based on measurements of phosphocreatine [13,14].

To help isolate such overlapping spectral lines, quantum filters and spectral editing pulse sequences can be used to eliminate background signals. While widely used in $^1$H NMR spectroscopy, there are few examples for $^{31}$P. Jayasunder *et al*. used a multi-quantum filter to remove interfering 2,3-diphosphoglycerate signal from measurements of inorganic phosphate [15], and Tsai *et al*. used double-quantum filtered HETCOR to study dentin with solid-state NMR [16]. Brindle *et al*. used spectral editing to suppress phosphomono- and phosphodiester signals [17].

Recently, a new class of quantum filters and spectral editing techniques have been developed based on the long-lived nuclear spin singlet state [18-23]. The singlet state can be created in pairs of coupled nuclear spins when the spins are in the magnetically equivalent or near-equivalent conditions. This condition can be satisfied when the chemical shift is small relative to scalar coupling, such that $\Delta\nu \ll J$, or through the application of decoupling via CW spin-locking or a pulse train. The resulting singlet states can potentially have long lifetimes far beyond $T_1$ and can also be isolated from other signals via appropriate pulse sequences. The structure of ADP and NAD, each with a pair of phosphate groups on which $^{31}$P-$^{31}$P singlet order can be prepared, appear to lend themselves to such a detection strategy (Fig. 1).

To test this method, we created singlet states in the $^{31}$P pairs of ADP and NAD$^+$. We measured the longitudinal relaxation time ($T_1$) and singlet relaxation time ($T_S$) under both neutral and basic conditions. We then used a singlet quantum filter via the SUCCESS sequence [18] to isolate the ADP and NAD$^+$ spectra in the presence of a large ATP background.

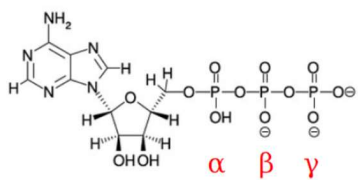
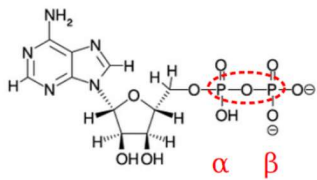
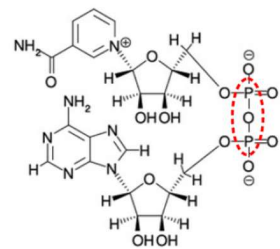

Figure 1: Structures of ATP, ADP, and NAD$^+$. Singlet order can be prepared on the $^{31}$P pairs of ADP and NAD$^+$ (circled). The triphosphate group of ATP does not support a singlet, because as a spin-0 state, the singlet must be composed of an even number of coupled spin-1/2 nuclei.

**Results**

$^{31}$P NMR spectra and measurements of T$_1$ and T$_S$ were acquired for ADP and NAD$^+$ (50 mM in pH 7.0 phosphate buffer) at 4.7 T (81 MHz $^{31}$P frequency). The spectrum of ADP consisted of two pairs of widely split lines typical of the weak coupling condition, with $\Delta \nu = 128.9$ Hz and $J_{PP} = 21.2$ Hz. NAD$^+$ exhibited a second-order pattern reflecting a much smaller frequency difference, with $\Delta \nu = 24.2$ Hz and $J_{PP} = 20.2$ Hz. T$_1$ times were between 2.5 and 2.9 s, while T$_S$ was ~200 ms for ADP and 470 ms for NAD$^+$ (Table 1), giving ratios $T_S/T_1 \approx 0.2$ and $T_S/T_1 \approx 0.5$ for ADP and NAD$^+$, respectively. The singlet lifetime is clearly not long-lived and is in fact significantly shorter than T$_1$, possibly due to strong relaxation from nearby proton spins interacting asymmetrically with the phosphorus nuclei [24]. At pH 7, the protons of both the ADP-β and the NAD$^+$ phosphate groups are fully dissociated, but one proton remains attached to the ADP-α phosphate group more strongly, likely driving $^{31}$P T$_1$ relaxation (pKa = 6.8) [25]. Therefore, as a test, a basic 50 mM ADP solution was prepared in 0.5 M sodium hydroxide solution. This increased the dissociation of the α phosphate's remaining proton, thereby increasing T$_1$ significantly and T$_S$ by a factor of eight. However, T$_S$/T$_1$ was still on the order of unity, indicating that other relaxation processes are also significant, likely from anti-correlated chemical shift anisotropy [26].

Table 1: Measured $^{31}$P longitudinal and singlet lifetimes for ADP and NAD$^+$ (50 mM) at 4.7 T (81 MHz $^{31}$P frequency).

| Molecule | T$_1$ (s) | T$_S$ (s) | T$_S$/T$_1$ |
|---|---|---|---|
| ADP-α neutral | 2.6 ± 0.2 | 0.49 ± 0.03 | 0.19 ± 0.02 |
| ADP-β neutral | 2.5 ± 0.1 | 0.53 ± 0.01 | 0.22 ± 0.01 |
| ADP-α basic | 3.9 ± 0.2 | 3.9 ± 0.7 | 1.0 ± 0.1 |
| ADP-β basic | 7.7 ± 0.3 | 4.2 ± 0.3 | 0.55 ± 0.04 |
| NAD$^+$ neutral | 2.9 ± 0.1 | 1.35 ± 0.01 | 0.47 ± 0.01 |

Next, we tested the $^{31}$P SUCCESS sequence on a mixture containing 30 mM ATP, 3 mM ADP, and 3 mM NAD$^+$ in a pH 7.0 phosphate buffer. Figure 1a shows a spectrum of the mixture measured with a 90°-FID sequence. Both ATP and ADP exhibit splitting patterns indicating weakly coupled $^{31}$P groups (additional ATP peaks at -20 ppm are not shown), and the ADP peaks are on the shoulders of the ATP peaks. One of the NAD$^+$ peaks lies directly beneath an ATP peak while the other is resolved. Figure 1b shows the results after using the SUCCESS sequence to target ADP. Approximately 39% of the ADP signal is retained, compared with only 2.6% of the ATP signal, resulting in a factor of 15-fold enhancement for ADP signal contrast. Figure 1c shows the results after targeting NAD$^+$ instead. Approximately 50% of the NAD$^+$ signal is retained, compared with only 1.2% of the ATP signal, resulting in a factor of 42-fold enhancement for NAD$^+$ signal contrast.

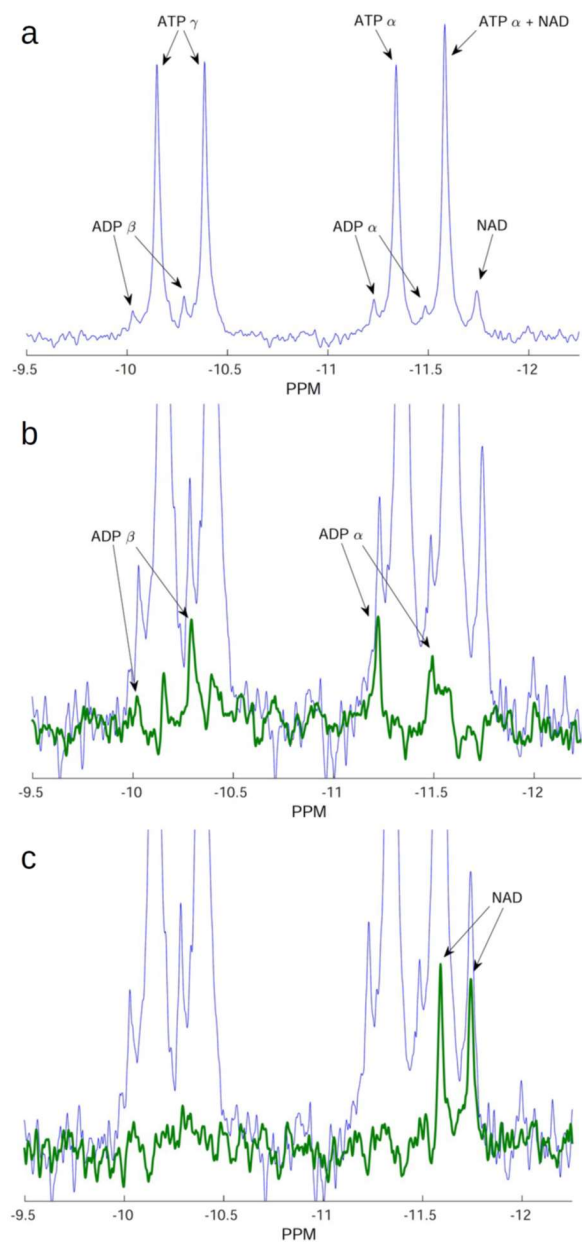

Figure 2. (a) $^{31}$P NMR spectrum of a mixture containing 30 mM ATP, 3 mM ADP, and 3 mM NAD$^+$ in a pH 7.0 phosphate buffer, measured with a 90°-FID sequence (2048 scans). ADP and NAD$^+$ signals are not fully resolved from ATP. (b) SUCCESS $^{31}$P NMR spectrum of the same sample targeting ADP eliminates 97% of the ATP signal, allowing three of the four ADP peaks to be resolved (8192 scans). (c) SUCCESS $^{31}$P NMR spectrum of the same sample targeting NAD$^+$ eliminates ATP and ADP signals, retaining only the NAD$^+$ doublet (8192 scans).

**Discussion and Conclusion**

Our results show that it is possible to create, and detect with NMR, nuclear spin singlet states on pairs of $^{31}$P nuclei in common biomolecules. However, these singlet states relax significantly faster than $T_1$ under typical conditions, similar to the results reported by Korenchan *et al.* in tetrabenzyl pyrophosphate [26]. Measurements under basic conditions show that nearby protons are partly to blame, but their removal does not necessarily lead to a long-lived state. The strongest relaxation mechanism is likely chemical-shift anisotropy (CSA), which is absent in $^1$H systems but is often strong in $^{13}$C, $^{15}$N, and $^{31}$P systems [24,26,27]. Long-lived singlets can still be created in $^{13}$C and $^{15}$N pairs when molecules are small and highly symmetric. While it is possible that $^{31}$P systems with the appropriate criteria also exist, rotation about the single bonds connecting phosphate groups makes it unlikely for this class of molecules.

Despite the short singlet lifetime, the SUCCESS sequence performed well at removing ATP background NMR signal and isolating the spectra of ADP and NAD$^+$. This method is potentially useful for *in vivo* MR spectroscopy at lower 1.5T and 3T magnetic field strengths typical of human and animal imaging, particularly when the broader lines *in vivo* lead to stronger spectral overlap. As a special case of zero-quantum filters, singlet filtration techniques such as SUCCESS and others are not significantly affected by $B_0$ inhomogeneity. While we based our technique on the Sarkar three-pulse sequence for singlet preparation and readout [28], other sequences such as M2S, SLIC, and APSOC might be more appropriate at lower magnetic field strengths, especially for NAD$^+$, whose spins enter the nearly-equivalent regime at 1.5 T [29-31]. In that regime, it might be possible to also eliminate spin-locking for singlet preservation, so that specific absorption rate (SAR) can be minimized. The SUCCESS sequence can also be improved by using gradients as part of the quantum filter for isolating singlet state, but these were not available on our spectrometer.

In conclusion, we found that $^{31}$P nuclear spin singlet states in ADP and NAD have relatively short lifetimes, but such states can nevertheless be used to isolate the NMR spectral signature of these molecules. This result suggests there is utility in exploring singlet states even in systems where short lifetimes might be expected due to strong chemical shift anisotropy or out-of-pair couplings. In particular, when spin relaxation properties are sensitive to parameters such as pH, the $T_s/T_1$ ratio might also provide useful information about the chemical environment.

**Methods**

NMR spectroscopy was performed on a Bruker DMX 200 MHz spectrometer with a $^1$H/X dual channel probe (81 MHz for $^{31}$P). Reagents were purchased from Sigma Aldrich (St. Louis, MO, USA). Conventional $^{31}$P spectra were acquired with a 90°-FID sequence, and $T_1$ was measured with a standard inversion recovery sequence. Chemical shifts were referenced to inorganic phosphate. $T_S$ was measured with the SUCCESS sequence (below) using a series of spin-lock nutation times ($\tau_4$). The $^{31}$P spin-lock nutation frequency was 615 Hz. In all cases $^1$H decoupling was applied during acquisition (CW, 150 Hz nutation frequency).

The SUCCESS technique has been described previously and is shown in Figure 3 [18,23]. In this pulse sequence, a singlet precursor state is prepared on a selected spin pair and CW decoupling is applied to create singlet order. A short sequence of hard pulses following a polyhedral phase cycle is then applied to create a quantum filter that passes only singlet order, which is then returned to transverse magnetization for readout. Pulse phases for $\phi_1$, $\phi_3$, $\phi_4$, and $\phi_{acq}$ of the polyhedral phase cycle are given in Table 2. $\phi_2$ was 0° throughout. Delay times for ADP were $\tau_1 = 12.02$ ms, $\tau_2 = 17.16$ ms, and $\tau_3 = 2.57$ ms. The center frequency was -10.74 ppm. For NAD$^+$, delays were $\tau_1 = 15.5$ ms, $\tau_2 = 16$ ms, and $\tau_3 = 12.5$ ms, and the center frequency was -11.675 ppm. For both molecules, the spin-locking time was $\tau_4 = 100$ ms and the spin-lock nutation frequency was $\nu_n = 615$ Hz. 2048 or 8192 averages were taken with a delay of 14 s between measurements. A large number of averages were used for the SUCCESS demonstration so that residual ATP signal could be quantified.

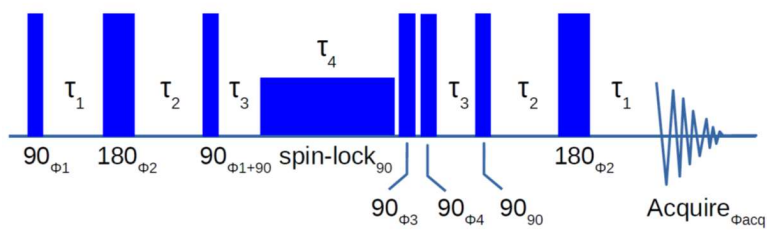

Figure 3. SUCCESS sequence for singlet state creation, filtration, and readout. Delays $\tau_1$, $\tau_2$, and $\tau_3$ are optimized to produce maximal singlet order on the target molecule and minimal singlet order in background molecules. Spin-locking is applied with a nutation frequency $\nu_n \geq 5\Delta\nu$, and spin-locking time $\tau_4$ can be varied to measure singlet lifetime or to create further contrast.

Table 2. Phase list for the SUCCESS sequence (in degrees).

| | |
|---|---|
| $\phi_1$ | 0,90,180,270,90,180,270,0,180,270,0,90,270,0,90,180,0,90,180,270,180,270,0,90 |
| $\phi_3$ | 0,0,0,0,90,90,90,90,180,180,180,180,270,270,270,270,0,0,0,0,180,180,180,180 |
| $\phi_4$ | 180,180,180,180,180,180,180,180,270,270,270,270,0,0,0,0,90,90,90,90,180,180,180,180 |
| $\phi_{acq}$ | 0,180,0,180,180,0,180,0,0,180,0,180,180,0,180,0,0,180,0,180,0,180,0,180 |

**Acknowledgments**

We acknowledge support from the NSF, Army, DARPA, and ARO MURI.

**Declaration of Competing Interests**

The authors share royalty interest in U.S. Patent 20150042331 covering the SUCCESS method.